\documentclass[12 pt]{article}
\usepackage{amssymb,amsthm,amsmath,color,graphics}
\theoremstyle{plain}

\theoremstyle{definition}
\newtheorem{definition}{Definition}

\newcommand{\ket}[1]{\ensuremath{\left|#1\right>}}

\newcommand{\ketbra}[2]{| #1 \rangle\langle #2 |}

\newcommand{\be}{\begin{equation}}
\newcommand{\ee}{\end{equation}}
\newcommand{\bc}{\begin{center}}
\newcommand{\ec}{\end{center}}
\newcommand{\bea}{\begin{eqnarray}}
\newcommand{\eea}{\end{eqnarray}}

\def\opone{\leavevmode\hbox{\small1\kern-3.8pt\normalsize1}}

\begin{document}

\bc {\LARGE \bf A Note on Quantum Separability}

\medskip
\renewcommand{\thefootnote}{\fnsymbol{footnote}}
{\footnotesize L. M. Ioannou$^{1,}$\footnote{lmi22@cam.ac.uk} and B.
C. Travaglione$^{1,2,}$\footnote{bct23@cam.ac.uk} \\ {\it $^1$Centre
for Quantum Computation, Department of Applied Mathematics
  and Theoretical Physics, \\ University of Cambridge, Wilberforce Road, Cambridge
  CB3 0WA, UK \\
$^2$Computer Laboratory, University of Cambridge, \\
William Gates Building, JJ Thomson Ave, Cambridge, CB3 0FD, UK }}
\ec

\bibliographystyle{unsrt}
\begin{quote}{\small This short note describes a method to tackle
the (bipartite) quantum separability problem.  The method can be
used for solving the separability problem in an experimental
setting as well as in the purely mathematical setting.  The idea
is to invoke the following characterization of entangled states: A
state is entangled if and only if there exists an entanglement
witness that detects it. The method is basically a search for an
entanglement witness that detects the given state.}\end{quote}

\section{Introduction}\label{sec_Introduction}
Entangled quantum states are interesting both from theoretical and
practical points of view.  Theoretically, entanglement is
connected with the confounding issue of nonlocality.  Practically,
entangled states are useful in quantum cryptography and other
quantum information processing tasks (for example, see \cite{NC00}
and references therein). The problem of determining whether a
state is entangled or separable is thus important and comes in two
flavors -- one mathematical, and the other experimental.

The mathematical problem of separability (for bipartite systems)
is defined as follows. Let $\mathcal{H}_{M,N}$ denote the set of
all Hermitian operators mapping $\mathbb{C}^M\otimes\mathbb{C}^N$
to $\mathbb{C}^M\otimes\mathbb{C}^N$. The set of bipartite
separable quantum states $S_{M,N}$ in $\mathcal{H}_{M,N}$ is
defined as the convex hull of the separable pure states
$\{\ketbra{\alpha}{\alpha}\otimes\ketbra{\beta}{\beta}\in
\mathcal{H}_{M,N}\}$, where $\ket{\alpha}$ is a norm-1 vector in
$\mathbb{C}^M$ and $\ket{\beta}$ is a norm-1 vector in
$\mathbb{C}^N$.  The set of separable states $S_{M,N}$ may be
viewed as a compact, convex subset of $\mathbb{R}^{M^2N^2}$ by
expressing each density operator as a real linear combination of
the canonical Hermitian generators of $SU(M)$ and $SU(N)$
\cite{TNWM02}. The quantum separability problem is now easily
defined as an instance of the Weak Membership problem
\cite{GLS88}: Given a convex set $K\subset\mathbb{R}^n$, a point
$p\in\mathbb{R}^n$, and an accuracy parameter $\delta>0$, assert
either that (i) $p\in S(K,\delta)$ (i.e. $p$ is ``almost in'' $K$)
or that (ii) $p\notin S(K,-\delta)$ (i.e. $p$ is ``not
$\delta$-deep within'' $K$), where $S(K,\delta)$ denotes the union
of all $\delta$-balls with centers belonging to $K$, and
$S(K,-\delta)$ denotes the union of all centers of all
$\delta$-balls contained in $K$ (in the standard Euclidean norm).
The separability problem has been shown to be NP-hard
\cite{Gur03}, thus any devised test for separability is likely to
require a number of computing resources that scales exponentially
with $M$ and $N$. There exist efficient ``one-sided'' tests for
separability, where the output of some polynomial-time computable
function of the matrix for $\rho$ can indicate that $\rho$ is
certainly entangled \cite{Per96,HH99,NK00,qphDPS03} or certainly
separable \cite{BCJLPS99,ZHSL98,GB02}, but not both. The algorithm
in \cite{qphDPS03} is in principle a one-sided test because it
requires an infinite amount of computational resources to detect
some entangled states.

The experimental flavor of the separability problem can be defined
as follows:  Given many physical copies of a completely unknown
quantum state $\rho\in\mathcal{H}_{M,N}$, determine whether $\rho$
is separable.  One way to solve this problem is to perform a full
state tomography in order to construct the density matrix for
$\rho$ to some precision $\delta>0$, and then solve the
mathematical separability problem.  If rather there is some
partial knowledge of $\rho$, then there are certainly more
options, such as testing for a violation of a specific Bell-type
inequality \cite{Bel64,CHSH69} or invoking entanglement witnesses
\cite{GHBELMS02,qphBMNMDM03}. As well, in the case where $MN\leq
6$, the positive partial transpose (PPT) test \cite{Per96,HHH96}
can be implemented physically \cite{qphHE01,qphCar03}, though
currently this approach is not experimentally viable.

In section \ref{sec_SolvQuSep} we describe a method to tackle the
mathematical separability problem in general. The idea is to invoke
the following characterization \cite{HHH96} of entangled states:  A
state $\rho$ is entangled if and only if there exists an entanglement
witness \cite{Ter00} that detects it. The method is basically a
search for an entanglement witness that detects the given state. In
section \ref{sec_UtilityExpSep} we describe how to use this method as
a novel tool for solving the experimental separability problem. We
conclude with a brief discussion, and point to some future directions
of research in section \ref{sec_Discussion}.

\section{Solving Quantum Separability}\label{sec_SolvQuSep}

In this note we use the following definition of ``entanglement
witness'', which differs slightly from the definition used in the
literature.
\begin{definition}
An \emph{entanglement witness} is any operator
$A\in\mathcal{H}_{M,N}$ for which there exists a state
$\rho\in\mathcal{H}_{M,N}$ such that
$$tr(A\sigma)<tr(A\rho) \quad \forall \ \sigma \in S_{M,N}.$$
\end{definition}
Recalling that $\mathcal{H}_{M,N}$ is isomorphic to
$\mathbb{R}^{M^2N^2},$ the above definition implies that for
entangled $\rho$ there exists a hyperplane in $\mathbb{R}^{M^2N^2}$
which separates $\rho$ from the set of all separable states
$S_{M,N}$. If we define the function
\begin{eqnarray}\label{eqn_bA}
b_A:= \max_{\sigma\in S_{M,N}}tr(A\sigma),
\end{eqnarray}
then the set $\lbrace x\in\mathcal{H}_{M,N}:\hspace{2mm}
tr(Ax)=b_A\rbrace$ is one such hyperplane. The function $b_A$ is
implicitly at the heart of the definition of ``entanglement
witness'' as it pins down which, if any, of the hyperplanes, with
normal A, separate the state $\rho$ from $S_{M,N}$. The hyperplane
defined by $A$ and $b_A$ is tangent to $S_{M,N}$ and is thus the
optimal hyperplane with normal vector $A$ that separates $\rho$
from $S_{M,N}$.

The function $b_A$ leads naturally to an algorithm for quantum
separability as follows. For $A$ such that $tr(A^2)=1$, define the
function $d_{A,\rho}$ as
\begin{eqnarray}\label{eqn_dA}
d_{A,\rho}:= b_A - tr(A\rho).
\end{eqnarray}
Geometrically, $d_{A,\rho}$ is the signed distance from the state
$\rho$ to the hyperplane defined by $b_A$. By using equations
\ref{eqn_bA} and \ref{eqn_dA}, and the definition of
``entanglement witness'', it follows that $\rho$ is entangled if
and only if
\begin{eqnarray}
\exists \ A  : \ d_{A,\rho}<0.
\end{eqnarray}

Thus we have reduced quantum separability to the task of finding
an $A$ such that $d_{A,\rho}$ is negative. By observing that the
ability to calculate $b_A$ gives an oracle for the Weak
Optimization problem, Theorem 4.4.7 from \cite{GLS88} shows that
it is possible to solve quantum separability with only
polynomially many evaluations of the function $b_A$. Thus, the
``hardness'' of quantum separability is contained in the
``hardness'' of evaluating $b_A$. In practice, and for
low-dimensional applications, well-known sophisticated techniques
(such as simulated annealing or interval analysis) for finding
global extrema would likely be sufficient to calculate $b_A$.
Unfortunately the algorithm derived in \cite{GLS88} is not
implementable on fixed precision computers, and it uses many more
evaluations of $b_A$ than is strictly necessary. In a later paper,
we hope to demonstrate a better algorithm which uses far fewer
evaluations of $b_A$ and thus, hopefully, is of practical use for
low-dimensional problems.

\section{Solving Separability with Partial Information}\label{sec_UtilityExpSep}

In this section, we briefly show how the above approach may be
used when only partial information about the state
$\rho\in\mathcal{H}_{M,N}$ is available. This is of particular use
in an experimental setting.

The state $\rho$ can be written
\begin{eqnarray}
\rho = \sum_{i=0}^{M^2-1}\sum_{j=0}^{N^2-1}
\rho_{ij}\lambda^M_i\otimes \lambda^N_j,
\end{eqnarray}
where $\rho_{ij}\in\mathbb{R}$ and the $\lambda^M_i$ and
$\lambda^N_j$ (as defined in \cite{TNWM02}) respectively generate the
special unitary groups $SU(M)$ and $SU(N)$. In this case, the
coefficients $\rho_{ij}$ are simply related to the expected values of
$\lambda^M_i\otimes \lambda^N_j$:
\begin{eqnarray}
\rho_{ij}=\langle\lambda^M_i\otimes
\lambda^N_j\rangle/4:=tr(\lambda^M_i\otimes \lambda^N_j\rho)/4.
\end{eqnarray}
Let $\Lambda:=\{\lambda^M_i\otimes
\lambda^N_j\}_{i=0,1,\ldots,M^2-1; j=0,1,\ldots,N^2-1}$. The
expected values of all elements of $\Lambda$ constitute complete
information about $\rho$.  Suppose only partial information about
$\rho$ has been obtained by an experimental procedure, that is,
only the expected values of the elements of a proper subset $T$ of
$\Lambda$ are known.

It helps to think of each density operator as a real vector of its
expected values.  With only $|T|$ expected values known, we now
effectively project all the density operators onto $\text{span}(T)$
by ignoring the components of the real vectors that correspond to the
unknown expected values of $\rho$.  Now each density operator,
including our unknown $\rho$, is represented by a point in a
$|T|$-dimensional ``expectation space''.  Note that the set of points
in this projective space representing all separable density operators
is still a convex set.  Call this convex set $\bar{S}_T$ and denote
its elements by $\bar{\sigma}\in\mathbb{R}^{|T|}$.  Similarly, let
$\bar{\rho}$ be the $|T|$-dimensional real vector of known expected
values of $\rho$. To represent the Hermitian operator $A=\sum_{X\in
T}a_XX$ in this space, we can use the real vector $\bar{A}$ of the
coefficients $a_X$ in order to have the correspondence
$tr(A\rho)=\bar{A}\cdot\bar{\rho}$, where ``$\cdot$'' denotes the
standard dot-product of two real vectors.

It is clear that the functions $b_A$ and $d_{A,\rho}$ in the previous
section can be redefined for the $|T|$-dimensional space:
\begin{eqnarray}
\bar{b}_{\bar{A}}:= \max_{\bar{\sigma}\in
\bar{S}_T}\bar{A}\cdot\bar{\sigma}
\end{eqnarray}
and
\begin{eqnarray}
\bar{d}_{\bar{A},\bar{\rho}}:= \bar{b}_{\bar{A}} -
\bar{A}\cdot\bar{\rho}.
\end{eqnarray}
If there exists $\bar{A}$ such that
$\bar{d}_{\bar{A},\bar{\rho}}<0$, then $\rho$ is entangled;
otherwise, more information is needed to determine the
separability of $\rho$. As expected values are being gathered
through experimental observation, they may be input to a computer
program that searches for $\bar{d}_{\bar{A},\bar{\rho}}<0$.
Finally, we point out that the idea of searching for an
entanglement witness in the span of operators whose expected
values are known was discovered independently and applied, in a
special case, to quantum cryptographic protocols in
\cite{qphCLL03}.

\section{Discussion}\label{sec_Discussion}

By looking at quantum separability as a mathematical problem in
the real Euclidean space $\mathbb{R}^{M^2N^2}$ and slightly
altering the definition of entanglement witness, we have show that
quantum separability can be solved in oracle-polynomial time, for
a rather natural looking oracle. This allows us to highlight the
fact that the ``hard'' part of quantum separability is contained
in the function $b_A$. The method we describe for solving quantum
separability also gives experimentalists a tool for potentially
determining if an unknown state is entangled by measuring only a
subset of the expected values which completely describe the state.
This method effectively trades quantum resources (additional
copies of $\rho$) for classical resources (a computer able to
calculate $\bar{b}_{\bar{A}}$). As well as providing a practical,
implementable algorithm for low-dimensional quantum separability,
some open questions which we hope to address in the future include
getting a tight (exponential) bound on the complexity of
calculating $b_A$, and determining the average number of expected
values required to detect a random unknown quantum state.

\section{Acknowledgements}
We would like to thank Carolina Moura Alves, Coralia Cartis, Donny
Cheung, Artur Ekert, and Tom Stace for useful discussions.  We
acknowledge support from the EU under project RESQ
(IST-2001-37559).  LMI also acknowledges support from CESG (UK)
and NSERC (Canada).  BCT also acknowledges support from CMI.


\end{document}